\documentclass[aps,prl,twocolumn,superscriptaddress,showpacs,amsmath,amssymb,longbibliography]{revtex4-1}
\usepackage[utf8]{inputenc}
\usepackage{graphicx,graphics}
\usepackage{hyperref}
\hypersetup{
  colorlinks=true,linkcolor=blue,urlcolor=blue,citecolor=blue}
\begin{document}
\title{Polarizing the Medium: Fermion-Mediated Interactions between Bosons}

\author{Dong-Chen Zheng}
\affiliation{Fujian Provincial Key Laboratory for Quantum Manipulation and New Energy Materials, College of Physics and Energy, Fujian Normal University, Fuzhou 350117, China}
\affiliation{Fujian Provincial Collaborative Innovation Center for Advanced High-Field Superconducting Materials and Engineering, Fuzhou, 350117, China}
\author{Lin Wen}
\affiliation{College of Physics and Electronic Engineering, Chongqing Normal University, Chongqing 401331, China}
\author{Chun-Rong Ye}
\affiliation{Fujian Provincial Key Laboratory for Quantum Manipulation and New Energy Materials, College of Physics and Energy, Fujian Normal University, Fuzhou 350117, China}
\affiliation{Fujian Provincial Collaborative Innovation Center for Advanced High-Field Superconducting Materials and Engineering, Fuzhou, 350117, China}
\author{Renyuan Liao}\email{ryliao@fjnu.edu.cn}
\affiliation{Fujian Provincial Key Laboratory for Quantum Manipulation and New Energy Materials, College of Physics and Energy, Fujian Normal University, Fuzhou 350117, China}
\affiliation{Fujian Provincial Collaborative Innovation Center for Advanced High-Field Superconducting Materials and Engineering, Fuzhou, 350117, China}
\date{\today}
\begin{abstract}
   We consider a homogeneous mixture of bosons and polarized fermions. We find that long-range and attractive fermion-mediated interactions between bosons have dramatic effects on the properties of the bosons. We construct the phase diagram spanned by boson-fermion mass ratio and boson-fermion scattering parameter. It consists of stable region of mixing and unstable region toward phase separation. In stable mixing phase, the collective long-wavelength excitations can either be well-behaved with infinite lifetime or be finite in lifetime suffered from the Landau damping. We examine the effects of the induced interaction on the properties of weakly interacting bosons. It turns out that the induced interaction not only enhances the repulsion between the bosons against collapse but also enhances the stability of the superfluid state by suppressing quantum depletion.
\end{abstract}
\maketitle
Ultracold atoms offer fascinating opportunities for investigating quantum many-body problems that relevant to fields as diverse as condensed matter physics, statistical physics, quantum chemistry, and high energy physics~\cite{GEO14,BLO17}. Of particular interest is Bose-Fermi mixtures~\cite{ONO16}, which allows one to explore the intriguing physics associated with the interplay between atoms of different quantum statistics. On the experimental side, tremendous progress have been achieved, which include controlling and characterizing the interspecies interactions~\cite{ING01,FER02,ROA02,KET04,JIN04,OSP06,BON06,CHI17,YOU20}, realizing mixture of Bose and Fermi superfluids~\cite{SAL14,SAL15,PAN16,GUP17,PAN18}, and probing physics of the phase separation state~\cite{GRI18,GRI19}. On the theoretical side, intense attentions have been paid to study ground-state properties~\cite{MOL98,SMI00,GIO02,ROT02,PAR08,WET11,ZHA14}, nature of excitations~\cite{HAN02,HUI03,TIM04,ZHA14,SAR15}, boson-mediated fermionic superfluidity~\cite{VIV00,STO00,SUZ08,BRU16,BRU18}, collective dynamics~\cite{SOR19,HUA20}, and  formation of exotic quantum phases~\cite{BLA03,CHI14,LIU15,STR16,CUI18,GAJ19,HAN19}.

Very recently, adding to the new excitements are the observations of fermion-mediated long-range interactions between bosons~\cite{CHI19,BRU19,OZE20}. The long-range nature of these mediated interactions enriches the toolbox for controlling coherent interactions~\cite{CHI10} and opens up the possibility of correlating distant atoms and preparing new quantum phases~\cite{BLA03,SAC08}. There have been theoretical attempts~\cite{TIM08,SPI14,BRU15} for understanding such fermion-mediated interactions based on the linear response theory. Given current experimental relevance, thorough theoretical understanding and identifying new features arising from fermion-mediated interactions becomes an urgent task.

In this work, we shall carry out a comprehensive study on the fermion-mediated interactions in Bose-Fermi mixtures, with the aim of laying down a solid and tractable framework to treat such problems, fully characterizing the mediated interactions, and elucidating the effects of the induced interactions on the bosons. First, we shall start from the functional representation of the partition function of the system. By tracing out the fermionic degrees of freedom, we obtain an effective action solely in terms of bosonic degrees of freedom, so that we can isolate the mediated effects of the fermions on the bosons. Second, we will examine the induced interaction at static limit in order to obtain an effective interaction potential. Third, we construct a phase diagram by taking account of both phase stability and the Landau damping of Bogoliubov excitations arising from density response from the Fermi gases. Finally, we examine the quantum fluctuations in the presence of the effective potential on the properties of the bosons.

We consider a homogeneous mixture of Bose gases and spin-polarized Fermi gases, described by the following grand canonical Hamiltonian
\begin{subequations}
\begin{eqnarray}
   H&=&H_B+H_F+H_{I},\\
   H_B&=&\int d\mathbf{r} \phi^\dagger(\mathbf{r}) \left(-\frac{\hbar^2\nabla^2}{2m_B}-\mu_B\right)\phi(\mathbf{r}),\\
   H_F&=&\int d\mathbf{r} \psi^\dagger(\mathbf{r}) \left(-\frac{\hbar^2\nabla^2}{2m_F}-\mu_F\right)\psi(\mathbf{r}),\\
   H_{I}&=&\!\int d\mathbf{r} \left(g_{I}\psi^\dagger\psi \phi^\dagger\phi+\frac{g}{2}\phi^\dagger\phi^\dagger\phi\phi\right).
\end{eqnarray}
\end{subequations}
 For bosons, $\phi(\mathbf{r})$ is the field operator, $m_B$ is the mass of an atom, and $\mu_B$ is the chemical potential. For fermions, $\psi(\mathbf{r})$ is the field operator, $m_F$ is the mass of an atom, and $\mu_F$ is the chemical potential. In the interaction term $H_I$, the coupling $g_{I}=2\pi\hbar^2 a_{FB}(m_F^{-1}+m_B^{-1})$ accounts for the interactions between the fermions and the bosons, and $g=4\pi\hbar^2 a_{BB}/m_{B}>0$ accounts for the repulsive interactions between bosons, where $a_{FB}$ and $a_{BB}$ are the corresponding s-wave scattering lengths. For convenience, we define the Fermi momentum $k_F=(6\pi^2n_F)^{1/3}$ with $n_F$ being the number density of Fermi gases, the Fermi velocity $v_F=\hbar k_F/m_F$ and the corresponding Fermi energy $E_F=\hbar^2k_F^2/2m_F$. We shall take natural units by setting $\hbar=k_B=1$ for sake of simplicity from now on.

Within the framework of the imaginary-time field integral, the partition function of the system can be cast as $\mathcal{Z}=\int d[\bar{\psi},\psi]d[\phi^*,\phi]e^{-S}$ with the action given by~\cite{SIM10} $S=\int_0^{\beta} d\tau \left[H+\int d^3\mathbf{r} (\bar{\psi}\partial_\tau\psi+\phi^*\partial_\tau\phi)\right]$, where $\beta=1/T$ is the inverse temperature. Carrying out the integration over the fermionic degrees of freedom, we obtain an effective action solely in terms of bosonic degrees of freedom $S_{eff}=S_{B}-Tr\ln\mathcal{M}$, where
\begin{eqnarray}
   \!S_{B}&=&\int d\tau d\mathbf{r} \left[\phi^*(\partial_\tau-\frac{\nabla^2}{2m_B}-\mu_B)\phi+\frac{g}{2}(\phi^*\phi)^2\right]\!,\nonumber\\
   \mathcal{M}&=&\partial_\tau-\frac{\nabla^2}{2m_F}-\mu_F+ g_I\phi^*\phi.
\end{eqnarray}
Up to this level, the formal manipulation of the partition function is exact. To distill low-energy physics, we shall resort to some sorts of approximations to be elaborated on.

 To proceed, we may write $\phi^*\phi=\rho_0+\sum_{q\neq 0}\rho_qe^{iqx}$, and we set $\mathcal{M}=-\mathcal{G}^{-1}+\mathcal{M}_1$ where   $\mathcal{G}^{-1}=-\partial_\tau+\nabla^2/2m_F+\mu_F-g_I\rho_0$ is the inverse fermionic Green's function and $\mathcal{M}_1=g_I\sum_{q\neq 0}\rho_qe^{iqx}$, with $x$ being space-time coordinate. This allows one to write
 $Tr\ln\mathcal{M}=Tr\ln(-\mathcal{G}^{-1})+Tr\ln{(1-\mathcal{GM}_1)}$ and to perform series expansions as follows
\begin{eqnarray}
   -Tr\ln(1-\mathcal{GM}_1)=\sum_{l=1}\frac{1}{l}Tr\left[(\mathcal{G}\mathcal{M}_1)^l\right].
   \label{eq:expansion}
\end{eqnarray}
To fully exploit the translational invariance of the system, we shall transform the above to momentum-frequency representation ($q\equiv(\mathbf{q},iw_m)$) resulting in
\begin{subequations}
\begin{eqnarray}
   &&Tr(\mathcal{GM}_1)=\mathcal{M}_1(0)\sum_{(\mathbf{k},iw_n)} \mathcal{G}(k)=0,
   \label{eq:one}\\
   &&\frac{1}{2}Tr\left[(\mathcal{GM}_1)^2\right]=\beta V\frac{g_{I}^2}{2}\sum_{q\neq 0}\Pi_q\rho_q\rho_{-q},
   \label{eq:two}\\
   &&\Pi_q=\frac{1}{\beta V}\sum_{(\mathbf{k},iw_n)}\mathcal{G}(k)\mathcal{G}(k+q).
   \label{eq:three}
\end{eqnarray}
\end{subequations}
Several comments are in order: For the expansion in Eq.~$(\ref{eq:expansion})$, the $l=1$ term vanishes, as can be seen from Eq.~$(\ref{eq:one})$; The $l=2$ term corresponds to induced two-body interactions between bosons, as can be seen from Eq.~$(\ref{eq:two})$ and Eq.~$(\ref{eq:three})$, where we have defined the so-called polarization function $\Pi_q$; $V$ is the volume of the system, $w_n=\pi(2n+1)/\beta$ is fermionic Matsubara frequencies, while $w_m=2\pi n/\beta$ is the bosonic Matsubara frequencies, where $n$'s are integers; We will neglect $l>=3$ terms, as they represent induced three-body or more than three-body  interactions for bosons, which are usually irrelevant for dilute gases. To be concrete, the effective action for the system is approximated as
\begin{eqnarray}
   S_{eff}=S_{B}-Tr\ln{(-\mathcal{G}^{-1})}+\beta V\frac{g_{I}^2}{2}\sum_{q\neq0}\Pi_q\rho_q\rho_{-q}.\label{eq:s}
\end{eqnarray}
We follow the standard Bogoliubov decomposition by splitting the bosonic field $\phi$ into a mean-field part $\phi_0$ and a fluctuating part $\varphi$: $\phi=\phi_0+\varphi$. By retaining the fluctuating fields up to the quadratic order, we approximate the effective action as $S_{eff}\approx S_0+S_g$, where $S_0$ is the mean-field action and $S_g$ is the gaussian action with quadratic orders of the fluctuating fields $\varphi_q^*$ and $\varphi_q$. Employing $\Omega=-\ln{\mathcal{Z}}/\beta V$, we obtain that the grand potential density at the mean-field level  $\Omega^{(0)}=S_0/\beta V$ becomes
\begin{eqnarray}
  \Omega^{(0)}&=&\frac{g|\phi_0|^4}{2}-\mu_B|\phi_0|^2-\frac{1}{\beta V}\sum_\mathbf{k}\ln{(1+e^{-\beta\xi_\mathbf{k}})},
\end{eqnarray}
where $\xi_\mathbf{k}=\mathbf{k}^2/2m_F-\mu_F+g_I|\phi_0|^2$.

Minimization of $\Omega^{(0)}$ with respect to the condensate order parameter $\phi_0^*$ leads to the Hugenholz-Pines theorem~\cite{HP59} determining the chemical potential $\mu_B=g_In_F+g|\phi_0|^2$. Without loss of generality, we shall take $\phi_0=\sqrt{n_{B}}$, where $n_B$ is the condensate density of the Bose gases. The self-consistent condition for the fermion density is determined via $n_F=-\partial\Omega^{(0)}/\partial\mu_F$, yielding
\begin{eqnarray}
    n_F=\frac{1}{V}\sum_\mathbf{k}f(\xi_\mathbf{k}),
\end{eqnarray}
where $f(x)=1/[1+\exp(\beta x)]$ is the Fermi-Dirac distribution function. Solving the above equation we obtain the chemical potential for the Fermi gases: $\mu_F=E_F+g_In_B$.

At the mean-field level, the ground-state energy density can be obtained via $E_G^{(0)}=\Omega^{(0)}+\mu_Fn_F+\mu_Bn_B$, yielding
\begin{eqnarray}
    E_G^{(0)}=\frac{3}{5}n_FE_F+\frac{g}{2}n_B^2+g_In_Fn_B.
\end{eqnarray}
For the system to be stable, we naturally require that the Hessian matrix $\partial^2E_G^{(0)}/\partial n_i\partial n_j$ ( $i,j=F,B$) constructed for the ground-state energy $E_G(n_F,n_B)$ to be positively definite, which leads to an upper bound for the fermion density
\begin{eqnarray}
   n_F^{1/3}<\frac{g}{3m_Fg_I^2}(6\pi^2)^{2/3},\label{eq:con}
\end{eqnarray}
which gives the condition for mechanical stability of the system~\cite{VIV00}.

As seen from the effective action in Eq.~$(\ref{eq:s})$, the Hamiltonian describing induced two-body interactions between bosons through coupling with fermions is given by
\begin{eqnarray}
  H_{ind}&=&\frac{g_I^2}{2}\sum_{\mathbf{q}\neq0}\sum_\mathbf{k,p}\Pi_\mathbf{q}\phi_{\mathbf{k+q}}^\dagger\phi_{\mathbf{p-q}}^\dagger\phi_{\mathbf{p}}\phi_{\mathbf{k}}.
\end{eqnarray}
Here, $\Pi_\mathbf{q}\equiv\Pi_{(\mathbf{q},0)}$ is the polarization function evaluated at the static  limit at zero temperature, which reads
\begin{eqnarray}
  \Pi_\mathbf{q}=-\frac{d(E_F)}{4}\left[1+\frac{k_F^2-q^2/4}{k_Fq}\ln\left\vert\frac{q+2k_F}{q-2k_F}\right\vert\right].\label{eq:pi}
\end{eqnarray}
where $d(E_F)=m_Fk_F/\pi^2$ is the density of states at the Fermi energy.

This corresponds to an induced pairwise interaction potential between two Bose atoms with relative coordinate  $\mathbf{r}$, given by
\begin{subequations}
\begin{eqnarray}
   V_{ind}(\mathbf{r})&=&-\frac{d(E_F)g_I^2}{4}V_{RKKY}(r),\\
   V_{RKKY}(r)&=&\frac{\sin{(2k_Fr)}-2k_Fr\cos{(2k_Fr)}}{2\pi k_Fr^4}.
\end{eqnarray}
\end{subequations}
The induced attractive long-range interaction is of the RKKY type~\cite{RKKY54} in real space, where it decays at $1/r^3$ and shows the Friedel oscillations at a wave vector $2k_F$.

The gaussian action for the bosonic fluctuating fields can be written as $S_g=\frac{1}{2}\sum_q \Phi_q^\dagger \mathcal{G}_B^{-1}(q)\Phi_q$ with $\Phi_q=(\varphi_q,\varphi_{-q}^*)^T$ and the matrix $\mathcal{G}_B^{-1}(q)$ given by
\begin{eqnarray}
  \mathcal{G}_B^{-1}(q)=\begin{pmatrix}-iw_m+\epsilon_\mathbf{q}+A_q &A_q\\
  A_q & iw_m+\epsilon_\mathbf{q}+A_q
  \end{pmatrix},
\end{eqnarray}
where  $\epsilon_\mathbf{q}=\mathbf{q}^2/2m_B$ and $A_q=(g+g_I^2\Pi_q)n_B$ . The quasiparticle spectrum $\omega(\mathbf{q})$ and the damping rate $\gamma(\mathbf{q})$  can be obtained by seeking solutions of the secular equation $det\mathcal{G}_B^{-1}(\mathbf{q},\omega-i\gamma)=0$ with substitution of $\Pi_q|iw_m\rightarrow \omega+i0^\dagger$. The real part of polarization function $\Pi_q$ determines the shift of the spectrum while the imaginary part gives rise to the damping of the excitations.
\begin{figure}[t]
\includegraphics[width=1.0\columnwidth]{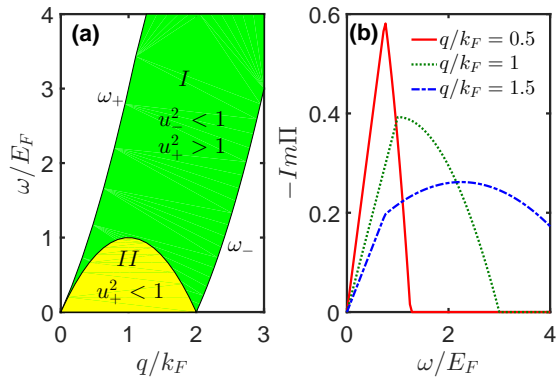}
\caption{(color online) (a) The shade region is the range where the imaginary part of the polarization function differs from zero, and it is referred as particle-hole continuum, since it is the region of single-particle excitations, whereby a particle below the Fermi surface is excited to above the Fermi surface. Outside this region, it is not possible to conserve energy and wave vector in a single-particle excitation process. (b) The imaginary part of the polarization function $Im\Pi(\mathbf{q},\omega)$ [in units of $d(E_F)$] as a function of frequencies $\omega$ at given different typical momentum amplitude $q$.}
\label{fig1}
\end{figure}

 By analytic continuation to real frequency($i\omega\rightarrow \omega+i0^\dagger$), one obtains the real part and the imaginary part of the polarization function, so-called the Lindhard function~\cite{LIN54}
\begin{widetext}
\begin{subequations}
\begin{eqnarray}
  Re\Pi(\mathbf{q},\omega)&=&-\frac{d(E_F)}{4}\left[1-\frac{1-u_-^2}{2q/k_F}\ln{\left\vert\frac{1+u_-}{1-u_-}\right\vert}+\frac{1-u_+^2}{2q/k_F}\ln{\left\vert\frac{1+u_+}{1-u_+}\right\vert}\right],\\
  Im\Pi(\mathbf{q},\omega)&=&-d(E_F)\frac{\pi k_F}{8q}\left[(1-u_-^2)\Theta(1-u_-^2)-(1-u_+^2)\Theta(1-u_+^2)\right],\label{eq:imp}
\end{eqnarray}
\end{subequations}
\end{widetext}
where $u_\pm=\omega/qv_F\pm q/2k_F$. Physically, it describes the response of the Fermi gases under external density perturbations exerted by the Bose gases. The imaginary part of the polarization function provides essential information for the damping of excitations, as it can be related to the dynamical structure factor through the fluctuation-dissipation theorem~\cite{COL15}. From Eq.~(\ref{eq:imp}), we find that it has contributions  from two situations enforced by Dirac-delta function: one is for $u_-^2<1$, and the other is from $u_+^2<1$. We show the region in momentum-frequency space where the imaginary part of the polarization function $Im\Pi(\mathbf{q},\omega)$ differs from zero in  Fig.~$\ref{fig1}(a)$. In the plot, region I is defined as $u_-^2<1$ and $u_+^2>1$, while region II satisfies $u_+^2<1$. The upper (lower) bound is given by $\omega_\pm/E_F=(q/k_F)^2\pm 2q/k_F$. In panel (b), we plot $Im\Pi$ as a function of frequencies $\omega$ for three typical momenta amplitude $q/k_F=0.5,1.0,1.5$. The curve will be linear when the corresponding ($\mathbf{q},\omega$) lying in region II, where both terms in Eq.~(\ref{eq:imp}) contribute. Otherwise, the curve will be parabolic when the corresponding ($\mathbf{q},\omega$) lying in region I, where only the first term in Eq.~(\ref{eq:imp}) contributes.
\begin{figure}[t]
\includegraphics[width=1.0\columnwidth]{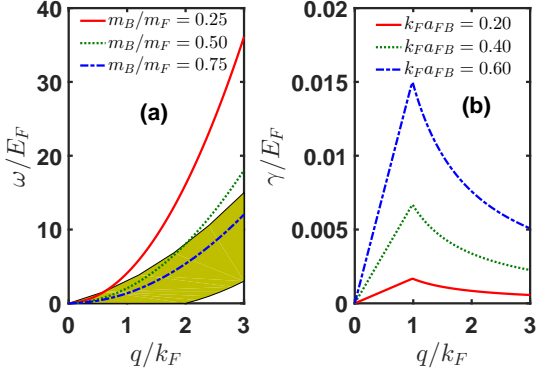}
\caption{(color online) Properties of the Bogoliugov quasiparticles: (a) the excitation energy $\omega/E_F$ where $k_Fa_{FB}=0.3$ and (b) the Landau damping rate $\gamma/E_F$ where $m_B/m_F=1$. The shade region is referred as particle-hole continuum. The quasi-particle spectrum laying outside of the shadow region is well-defined, being immune from the Landau damping. We set $k_Fa_{BB}=0.3$ and $n_B/n_F=0.2$. }
\label{fig2}
\end{figure}

With the information of the polarization function at hand, it is straightforward to obtain the quasiparticle spectrum $\omega(\mathbf{q})$ and the damping rate $\gamma(\mathbf{q})$, shown in Fig.~\ref{fig2}. From Eq.~(\ref{eq:imp}), one obtains that the region for damping to occur is given by the inequality constraint $(q/k_F)^2-2q/k_F< \omega/E_F<(q/k_F)^2+2q/k_F$, as shown in the shade region on panel (a) in Fig.~\ref{fig2}. At small momenta, the spectrum is phonon-like with the sound velocity given by $c=\sqrt{(g+g_I^2\Pi_0)n_B/m_B}$. The positivity of the sound velocity yields the stability constraint $(k_Fa_{FB})^2<2\pi k_Fa_{BB}m_Fm_B/(m_F+m_B)^2$, coincident to the mechanical stability provided in Eq.~(\ref{eq:con}). The mass of bosons $m_B$ has dramatic effects on the spectrum: at low momenta, the slope is inversely proportional to $\sqrt{m_B}$; while at  high momenta, it gives the mass for free particle. For sufficiently small $m_B/m_F$, the excitations can achieve infinite lifetime.  The damping of the excitations show sharp peak at wave-vector $q=k_F$ for all three typical boson-fermion scattering parameter $k_Fa_{FB}$.

\begin{figure}[t]
\includegraphics[width=1.0\columnwidth]{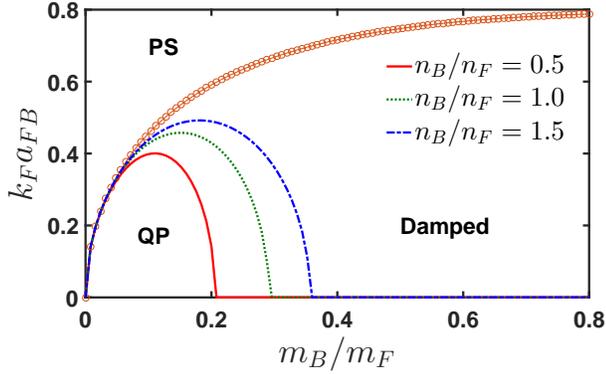}
\caption{(color online) Phase diagram spanned by mass ratio $m_B/m_F$ and interspecies coupling strength $k_Fa_{FB}$. PS stands for phase separation and QP stands for quasi-particle with infinite lifetime. Here we set $k_Fa_{BB}=0.4$, which sets bosons in a weakly-interacting regime. }
\label{fig3}
\end{figure}

We are now in a position to construct a phase diagram for the system. The stability constraint marks the transition line between stable mixing phase and phase separation (PS) into fermions and bosons~\cite{MOL98,SMI00,ROT02}, shown in Fig.~\ref{fig3}, which stays intact for different number density ratio. In the stable region, we can further classify it into quasiparticle excitations with infinite lifetime and  with finite lifetime due to the Landau damping, by which a particle absorbs an excitation of  momentum $\hbar \mathbf{q}$ and energy $\hbar\omega$ to allow it to move from beneath to above the Fermi surface, creating a particle-hole pair. To search for well-defined, long-lived excitations, we consider the region where $Im\Pi(\mathbf{q},\omega)=0$. This occurs when $\omega/qv_F>1+q/2k_F$ [see Eq.~(\ref{eq:imp})]. At long-wavelength, this becomes $c/E_F>2/k_F$, yielding
\begin{equation}
    \left(k_Fa_{FB}\right)^2\!<\!\frac{2\pi m_Fm_B}{(m_F+m_B)^2}\!\left(\! k_Fa_{BB}-\frac{3\pi}{2}\frac{n_F}{n_B}\frac{m_B^2}{m_F^2}\!\right).
\end{equation}
The phase diagram constructed is shown in Fig.~\ref{fig3}. The region of the quasiparticle excitations with infinite lifetime (QP) gets expanded by tuning up the number density ration $n_B/n_F$. It should be pointed out that we focus on the Landau damping of the collective long-wavelength excitation, where Beliaev damping is strongly suppressed at low momenta~\cite{LIU97}.

To examine the effects of the effective potential upon the Bose gases, we shall evaluate the ground-state energy correction arising from quantum fluctuations. The fluctuation correction to the thermodynamic potential is given by $\Omega_f=\frac{\beta}{2}Tr\ln \mathcal{G}_B^{-1}-\sum_\mathbf{q}(\epsilon_\mathbf{q}+A_\mathbf{q})$. At zero temperature, the corresponding ground-state energy correction, becomes renormalized as
\begin{eqnarray}
  \Delta E_G=\frac{1}{2}\sum_\mathbf{q}\left(\omega_\mathbf{q}-\epsilon_\mathbf{q}-A_\mathbf{q}+\frac{g^2n_B^2}{2\epsilon_\mathbf{q}}\right),
\end{eqnarray}
where the Bogoliugov spectrum is given by $\omega_\mathbf{q}=\sqrt{\epsilon_\mathbf{q}(\epsilon_\mathbf{q}+A_\mathbf{q})}$, and $A_\mathbf{q}=(g+g_I^2\Pi_\mathbf{q})n_B$.

\begin{figure}[t]
\includegraphics[width=1.0\columnwidth]{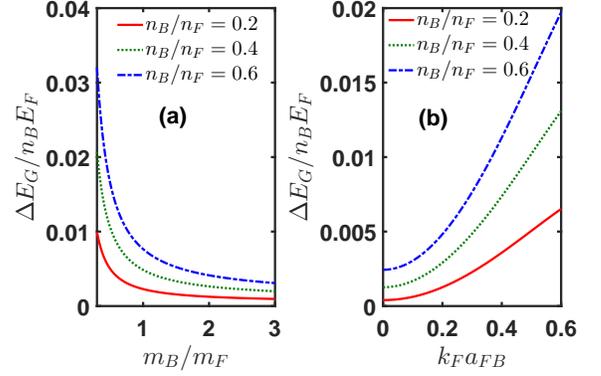}
\caption{(color online) Correction to the ground-state energy per density  $\Delta E_G/n_BE_F$ (a) as a function of mass ratio $m_B/m_F$ where $k_Fa_{FB}=0.3$ and (b) as a function of boson-fermion scattering length $k_Fa_{FB}$ where $m_B/m_F=1$. We set $k_Fa_{BB}=0.3$.}
\label{fig4}
\end{figure}
The behavior of  fluctuation correction to the ground-state energy $\Delta E_G$ is shown in Fig.~\ref{fig4}. In panel (a), we find that as the mass ratio $m_B/m_F$ increases the energy correction decreases, which is reasonable since the kinetic energy is inversely proportional to the mass of the bosons. It is interesting to notice that increasing the density ratio $n_B/n_F$ actually contributes to the enhancement of the energy correction. Shown in panel (b), the energy correction increases monotonically with boson-fermion coupling strength $k_Fa_{FB}$. For vanishing boson-fermion interaction, one can verify that  the energy correction recovers the Lee-Huang-Yang correction~\cite{LHY57} to the spinless weakling interacting bosons $\Delta E_G/(gn_B^2)=64/(15\sqrt{\pi})\sqrt{n_Ba_{BB}^3}$. It is remarkable that the correction of energy increases steadily with increasing $k_Fa_{FB}$. This raises the possibility of realizing quantum droplets states with enhanced quantum repulsion again collapse.

We turn to the depletion of the condensates due to quantum fluctuations, which provides key information about the robustness of the superfluid state. The number of excited particles is evaluated as
\begin{eqnarray}
    n_{ex}=\sum_q {\mathcal{G}_{B}}_{11}(\mathbf{q},iw_n)=\sum_\mathbf{q}\frac{\epsilon_\mathbf{q}+A_\mathbf{q}-E_\mathbf{q}}{2E_\mathbf{q}}.
\end{eqnarray}
\begin{figure}[t]
\includegraphics[width=1.0\columnwidth]{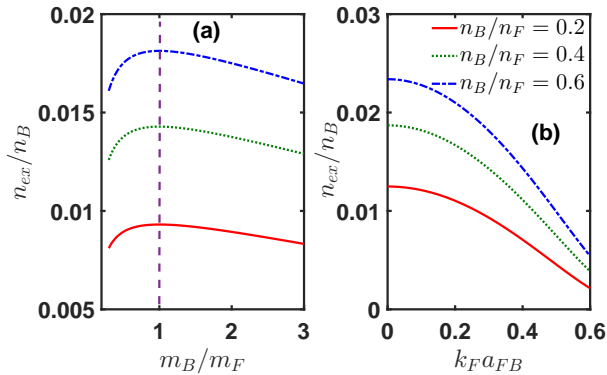}
\caption{(color online) Quantum depletion of the condensates $n_{ex}/n_B$ (a) as a function of mass ratio $m_B/m_F$ where $k_Fa_{FB}=0.3$ and (b) as a function of boson-fermion scattering parameter $k_Fa_{FB}$ where $m_B/m_F=1$. The vertical dash line intercepts with maxima of the curves at $m_B=m_F$. We set $k_Fa_{BB}=0.3$.}
\label{fig5}
\end{figure}
The variation of quantum depletion with respect to tuning parameters $m_B/m_F$ and $k_Fa_{FB}$ are shown in Fig.~\ref{fig5} for fixed density ratios. Aesthetically appealingly, $n_{ex}/n_B$ develops a maximum at equal mass $m_B=m_F$. Remarkably, increasing the boson-fermion interaction suppresses the quantum depletion, due to attractive nature of induced interaction between bosons. At zero boson-fermion coupling, it recovers the known result~\cite{NN99} for spinless weakly interacting Bosons $n_{ex}=(gn_B)^{3/2}/(3\pi^2)$.

In summary, we find that the induced interaction mediated by fermions between bosons are long-range attractive interactions, tunable with fermion density as well as boson-fermion scattering length. We map out the phase boundary separating stable region of mixing phases and unstable region toward phase separation. We show that the stable region can be further classified by damping of the excitations. The predicted damping rate can be probed experimentally via two-phonon Bragg spectroscopy~\cite{DAV05}. Extension of current work to trapped cases~\cite{ONO02,ONO16,ONO20} will facilitate the experimental verification of our predictions. Finally, we analyze the effects of the induced interactions on the ground-state energy correction and quantum depletion of the system. It suggests that by coupling to Fermi gases, weakling interacting bosons may form quantum droplet states~\cite{CUI18,GAJ19} with enhanced stability. We expect our study contribute to a better understanding of emergent phenomena associated with fermion-mediated interactions.

\section*{acknowledgments}
R.L. acknowledges funding from the NSFC under Grant No.11674058 and NCET-13-0734. L.W. was supported  by NSFC under Grant No. 11875010,
and by the Natural Science Foundation of Chongqing under Grant No. cstc2019jcyj-msxmX0217.

\end{document}